\documentclass{article}

\usepackage{graphicx}% Include figure files
\usepackage{dcolumn}% Align table columns on decimal point
\usepackage{bm}% bold math

\RequirePackage{xspace}
\usepackage{relsize}

\def\Kbar  {\kern 0.2em\overline{\kern -0.2em K}{}\xspace}

\def\Bbar    {\kern 0.18em\overline{\kern -0.18em B}{}\xspace}

\def\Qbar    {\kern 0.08em\overline{\kern -0.08em Q}{}\xspace}

\newcommand{\mev}{\ensuremath{\mathrm{\,Me\kern -0.1em V}}\xspace}
\newcommand{\mevc}{\ensuremath{{\mathrm{\,Me\kern -0.1em V\!/}c}}\xspace}
\newcommand{\mevcc}{\ensuremath{{\mathrm{\,Me\kern -0.1em V\!/}c^2}}\xspace}
\newcommand{\gev}{\ensuremath{\mathrm{\,Ge\kern -0.1em V}}\xspace}
\newcommand{\gevc}{\ensuremath{{\mathrm{\,Ge\kern -0.1em V\!/}c}}\xspace}
\newcommand{\gevcnospace}{\ensuremath{{\mathrm{\,Ge\kern -0.1em V\!/}c}}}
\newcommand{\gevcc}{\ensuremath{{\mathrm{\,Ge\kern -0.1em V\!/}c^2}}\xspace}
\newcommand{\be}{\begin{equation}}
\newcommand{\ee}{\end{equation}}
\newcommand{\benn}{\begin{equation*}}
\newcommand{\eenn}{\end{equation*}}
\newcommand{\bea}{\begin{eqnarray}}
\newcommand{\eea}{\end{eqnarray}}

\begin{document}
%\pagewiselinenumbers
% the following line is for submission
%\hspace{5.2in} \mbox{Fermilab-Pub-04/xxx-E}
%\bibliographystyle{apsrev}

\title{Revisiting Constraints on Fourth Generation Neutrino Masses}
%\title{{Search for fermion-pair decays}}
%\\ \vspace*{2.0cm}}

\author{L.M.~Carpenter
and A.~Rajaraman\\
{\small {\it University of California, Irvine, Irvine, California 92697}}}

\date{}
\maketitle
\begin{abstract}
We revisit the current experimental bounds on
fourth-generation Majorana neutrino masses, including the effects of right handed neutrinos.
Current bounds from LEPII are significantly altered by a global analysis. We show that the current bounds on fourth generation neutrinos decaying to eW ar $\mu$W can be reduced to about 80 GeV (from the current bound of 90 GeV), while a neutrino
 decaying  to $\tau$W
 %or into flavor democratic final states
 can be as light as 62.1 GeV.  The weakened bound opens up a neutrino decay channel for intermediate mass Higgs, and interesting multi-particle final states for Higgs and fourth generation lepton decays.
\end{abstract}

% 12.60.-i Models beyond the standard model
% 13.85.Rm Limits on production of particles
% 14.80.-j Other particles (including hypothetical)

%\pacs{12.60.-i, 13.85.Rm, 14.80.-j}

%\date{\today}

\section{Introduction}

One of the most natural possibilities for an extension of the standard model   is a
fourth copy of the three known generations of particles.
These fourth generation quarks and leptons  could not be very heavy, since
they acquire mass from chiral symmetry breaking.
They would thus be accessible at the LHC, with striking signatures even in early data.

While previous studies had argued that such fourth generation particles were incompatible
with electroweak precision data \cite{Amsler:2008zzb},  recent
work~\cite{Kribs:2007nz} has shown
that the existence of weak scale charged and neutral leptons is
allowed by electroweak precision data for heavier Higgs masses, with appropriate mass
splittings for the new particles. Furthermore,  certain anomalies
in the b-quark
sector can be ameliorated by a fourth
generation~\cite{Aaltonen:2007he,:2008fj,Hou:2005hd,Hou:2005yb,Soni:2008bc,:2008zza}.
This has led to a revival of interest in this
possibility (see~\cite{Holdom:2009rf} for a review).

In light of upcoming LHC searches, it is of particular
importance to understand the allowed parameter space of the fourth generation.
Several experiments constrain the parameter space of  fourth generation quarks.
Experimental measurements at the Tevatron
have set
limits of 311 GeV for the $t'$~\cite{Cox:2009mx}, and 338 GeV  for the $b'$~\cite{Aaltonen:2009nr}.
Many studies have also been done of the possibility of discovering fourth generation quarks at the LHC
\cite{Ozcan:2008zz,Cakir:2008su,Holdom:2007ap,Han:2006ip,Atre:2009rg,delAguila:2008ir,delAguila:2007em,Hung:2007ak,CuhadarDonszelmann:2008jp}.
 From these studies,  it appears that the LHC can discover or exclude fourth generation quarks to about a TeV.

On the other hand, constraints on the lepton sector of the new generation have not been studied
in as much detail. While LEP II has placed bounds on fourth generation neutrino masses, there has as yet been no search performed at the Tevatron (Tevatron sensitivity studies for fourth generation neutrinos were performed in~\cite{Rajaraman:2010wk}). This is a surprising omission, since fourth generation leptons are expected
to be lighter than fourth generation quarks, if the first three generations are any indication.
Indeed, the fourth generation neutrino should be considerably lighter than the quarks. The
fourth generation neutrino search is therefore particularly appealing.

Furthermore, the lepton sector is expected to be extremely rich in phenomenology. The reason is
that the relatively high mass scale for the neutrino poses a puzzle for building models of the fourth
generation. If the neutrino mass is generated by the dimension 5 operator  $\nu\nu HH\over M$,
then the suppression scale M cannot be too high (in this case, less than a few TeV), and there should be new particles
at this scale. The most natural assumption is that the right handed neutrino for the fourth
generation is not very heavy, and that the neutrino masses are generated by the analogue of
the seesaw mechanism, except that the scale of suppression is much lower. This  provides a reason
why the fourth generation neutrino is so much heavier
than the others. It also immediately implies that any phenomenological analysis should  include both the
left- and right-handed neutrino. In addition the charged lepton can also potentially play a role in the phenomenology of this sector.

In this note, we will revisit the LEPII bounds on neutrino masses, taking into
account the existence of both left- and right-handed neutrinos. We will not include the charged lepton in this analysis; we will return to this in future work. We find that
current bounds are significantly
diluted once this extra state is included. This has important consequences for future
collider searches, as the parameter space is enlarged considerably, with new interesting signals
that were not analyzed previously.

We will analyze the two-neutrino parameter space in more detail below. In section 2 we review the theory of fourth generation neutrinos, as well as their
production and decay.
In section 3, we review the LEP experimental searches for neutrinos, and find the efficiency of this search when applied to the more general parameter space.  We then analyze the
bounds imposed on the two-neutrino parameter
space from these searches and show that the parameter space can be considerably enlarged. Finally, we conclude with a discussion of future directions, in particular effects on Higgs searches.

\begin{table}[t]
\begin{center}
\begin{tabular}{|c|c|c|c||c|c|c|}
\hline
 &\multicolumn{3}{c||}{
 %Electron, Muon
 $ e, \mu$ mode} & \multicolumn{3}{c|}{$\tau$ mode}\\
\hline
$N_1$ mass & $\epsilon_{11}$ & $\epsilon_{12}$ & $\epsilon_{22}$ & $\epsilon_{11}$ & $\epsilon_{12}$ & $\epsilon_{22}$ \\
\hline
45 & .162 & .313 & .331  & .121 & .149 & .181\\
\hline
55  & .188 & .336 & .338 & .125 & .151 & .188  \\
\hline
65  & .224 & .342 & .384  & .110 & .147 & .196\\
\hline
75  & .251 & .342 & .369 & .114 & .149 & .199  \\
\hline
85  & .234 & .325 & .352 & .129 & .155 & .195 \\
\hline
\end{tabular}
\caption{Search efficiencies for $N_1N_1, N_1N_2$ and $N_2N_2$ processes respectively where $N_1$ decays
to eW, $\mu$W or $\tau$W.}
\label{tab:Eff}
\end{center}
\end{table}

\section{Review of Fourth Generation Neutrinos}
We will be following the notation of \cite{Katsuki:1994as}.

We consider an extension to the standard model by a fourth generation
of fermions and a right-handed neutrino.  The mass term for the neutrinos can be written as
\begin{eqnarray}
L_m=-{1\over 2}\overline{(Q_R^c
N_R^c)}\left(\begin{array}{cc}0&m_D\\m_d&M\end{array}\right)
\left(\begin{array}{c}Q_R\\ N_R\end{array}\right)+h.c.
\end{eqnarray}
where $\psi^c=-i\gamma^2\psi^*$.
This theory has two mass eigenvalues
\bea
 \nonumber m_1=-(M/2)+ \sqrt{m_D^2+{M^2/4}} \\ m_2=-(M/2)- \sqrt{m_D^2+{M^2/4}}
\eea

\noindent
with the corresponding eigenstates
 \bea
 N_1=\cos\theta Q_L^c+s_\theta N_R+\cos\theta  Q_L+s_\theta N_R^c
 \\
 N_2=-is_\theta Q_L^c+i\cos\theta  N_R+is_\theta Q_L-i\cos\theta  N_R^c
  \eea
  where we have defined the mixing angle
\bea
\nonumber \tan\theta=m_1/m_D\label{mixingangle}
\eea
Note in particular that $\theta=\pi/4$ corresponds to a pure Dirac state, while
$\theta=\pi/2$ corresponds to a pure Majorana state (the other fermion decouples in this limit).

In addition, there is a mass term for the fourth
generation lepton. We will assume that the lepton is heavier than
the two neutrinos; we will therefore not include it in our analysis.

The  neutrinos couple to the gauge bosons through the interaction term
$L=gW_\mu^+ J^{\mu +}
 +gW_\mu^- J^{\mu -}+gZ_\mu J^{\mu}$ where
\bea
 J^{\mu }
= {1\over 2\cos\theta_W}(-c^2_\theta\bar N_1\gamma^\mu \gamma^5N_1
-2is_\theta c_\theta\bar N_1\gamma^\mu N_2-s^2_\theta\bar N_2 \gamma^\mu\gamma^5N_2))
\\
J^{\mu +}=%{1\over {2}}
c_i\overline{(c_\theta N_1-i s_\theta N_2)}\gamma^\mu l^i_L~~~~~~~~~~~~~~~~~~~
\eea
where $c_i$ are analogous to the CKM matrix elements.

We now consider the possible decay modes of $N_1,N_2$. Since we have assumed that the fourth
generation lepton is heavy, the lighter neutrino $N_1$ can only decay through
a charged current interaction to $lW$, where $l$
is a lepton of the first  three generations. %=(e,\mu\tau)$ i
The relative branching ratios are set by the unknown $c_i$, and we
will  have to consider each possibility separately.

$N_2$, on the other hand, can decay either to $lW$ or to $N_1Z$.
The first decay mode is suppressed by the small mixing between the
fourth generation and the other three generations (which we shall assume to be much
smaller than the electroweak coupling). For most
masses, the second decay mode will dominate.
When the mass difference between the two neutrinos goes
to zero (the pseudo-Dirac limit), there is a phase space suppression
of the second mode.
We will assume that we do not have this extreme degeneracy and that
the mode $N_2 \rightarrow {N_1} $  dominates. We shall impose this by assuming that the
$m_2-m_1>10$ GeV.

Note also that in the Dirac limit, only the CKM suppressed
decay is allowed to occur, and the interference between the
various contributions kills the same sign dilepton decays.
This is expected since the Dirac fermion conserves fermion number.
Since we are assuming that the decay $N_2 \rightarrow N_1 Z$  dominates, this
interference does not occur. We therefore get same sign
dilepton decays for all the parameter space we consider.

\section{ LEP Constraints}

The existing constraints on fourth generation
neutrino masses mainly come from LEP II
~\cite{Achard:2001qw}.  These searches assumed that there was a single neutrino, which
was either Majorana or Dirac,
which decayed through a W boson, $N \rightarrow W^{+} l^{-} $.
The analysis depended on whether the lepton was e, $\mu$ or $\tau$.
If the
neutrino decayed to eW or $\mu$W, the events were required to satisfy the following requirements:

1) Two isolated leptons (same flavor) with a total energy less than 0.7 $E_{beam}$

2) Number of jets plus isolated leptons is at least 3

3) Hadronic energy exceeds 60 GeV and charged track multiplicity larger than 3

If the neutrino decayed to $\tau$W, the event selection depended on whether the tau decayed leptonically or hadronically. For leptonic decays, the events  had to pass the event selection above
with the same flavor requirement relaxed. If at least one tau decayed hadronically, the event was required to satisfy

1) Number of jets plus isolated leptons is at least 4

2) Polar angle of missing momentum in the range $25^0 < \theta_{miss} < 155^0$

3) Fraction of visible energy in the forward backward region ($20^0 > \theta$ or  $\theta> 160^0$)
should be less than 40\%.

4) All electron and muon energies less than 50 GeV.

5) Angle between most isolated track and track nearest to it should be greater than $50^0$, or
the angle between second most isolated track and track nearest to it should be greater than $25^0$.

6) Transverse momenta of two most isolated tracks should be greater than 1.2 GeV, and at least
one track must have a transverse momentum greater than 2.5 GeV.

For decay
channels where the  lepton is entirely e, entirely $\mu$ or entirely $\tau$,
mass exclusions were made at 90.7, 89.5, and 80.5 GeV respectively for
Majorana particles and 101.5, 101.3 and  90.3 GeV respectively for Dirac particles
(the Majorana mass bounds are lower then the Dirac bounds because Majorana fermion production
is accompanied by an extra velocity factor in the production cross section.) Similarly,
stable fourth generation neutrinos need to be fairly heavy (at least 40 GeV) to
escape constraints posed by the invisible width of the Z.

\begin{table}[t]
\begin{center}
\begin{tabular}{|c|c|c|c|c|c|c|}
\hline
CM Energy (GeV) &   192 &  196 &  200 &  202 &  205&  207\\
\hline
Luminosities  &  26 & 76  &  83 &  41 &  83 &  140 \\
\hline
\end{tabular}
\caption{Luminosities in pb$^{-1}$ at LEP II as a function of energy.}
\end{center}
\end{table}
We now consider the two neutrino parameter space.
%, and analyze constraints in the
%$m_1-m_2$ plane.
At LEP, the two neutrinos are produced through the Z by the processes
\bea
ee\rightarrow Z \rightarrow N_1N_1\rightarrow lWlW~~~~\nonumber
\\
ee\rightarrow Z \rightarrow N_1N_2\rightarrow lWlWZ~~\nonumber
\\
ee\rightarrow Z \rightarrow N_2N_2\rightarrow lWlWZZ\nonumber
%\left{\begin{array}{c}{
%N_1N_1,N_1N_2,N_2N_2
%}\end{array}   \right.
\eea
To calculate the new constraints on this parameter space, we need to find the sensitivity of the LEP analysis
to $N_1N_2$ and $N_2N_2$ production.

%We have reproduced their efficiencies by
 To find the efficiencies, we generated $ee\rightarrow N_iN_j$ events using MADGRAPH 4.4.32~\cite{Alwall:2007st}.
 The neutrinos were then decayed
 using BRIDGE~\cite{Meade:2007js}, and the events were hadronized using Pythia~\cite{pythia}.
 The efficiencies of the processes were calculated by simulating events and
examining how many passed the cuts described above.
For the case of $N_1N_1$ decaying to electrons and muons, we were able to reproduce the efficiencies
obtained by LEP; in particular, we obtain the same mass bound on the neutrinos
in the Majorana limit.  For the tau case, we had to scale our efficiencies by a factor of 1.3 to
obtain the LEP efficiency (to reproduce the mass bound in the Majorana limit).

For the $N_1N_2$ and $N_2N_2$ processes, we then apply the same scaling; viz.
we  scaled all the processes with tau final states by
a factor 1.3 to obtain our final efficiencies. We found that these efficiencies are
almost independent of the mass of $N_2$. The final  scaled efficiencies are shown in Table 1.

We note that in these analyses,
 we have assumed that the state $N_1$ decays entirely into a single
 species of lepton.  In
 any more general situation, the lower bound for all efficiencies is set
 by the tau search, since in the leptonic decay mode, the tau search uses the same search
 parameters as the electron and muon final state search.

The largest factor contributing to the features of the efficiencies was the existence of a hard well isolated final state lepton.  This causes the efficiencies for the detection of $N_1 N_2 $ and $N_2 N_2$ processes to be higher that that of $N_1 N_1$, as the decay of the heavy to light neutrino proceeds through a Z boson, which may produce additional final state leptons.  We  must note, however, that this will only be the case as long as the mass splitting between the two neutrino states is large enough.  As the neutrino masses approach each other, the final state  leptons from the offshell $N_2$ decay become soft, the isolated leptons are lost, and the detection efficiency drops.  If the mass difference is very small, $N_2$ lives long enough to decay outside of the vertex detector and the efficiency for heavy neutrino detection drops precipitously.  We have only explored the regions in the $m_1 m_2$ mass plan where the mass neutrino mass difference is greater than 10 GeV, and the search efficiency for $N_2$ remains high.  However we would expect that the least stringent mass bounds on neutrinos would actually come from the (possibly very fine-tuned) region where the neutrino  mass splitting is very small.

%\section{Revised Constraints}

We can now calculate the number of expected events at LEP. From the period 1999-2000, LEP collected 450 pb$^{-1}$ of data between 192-207 GeV~\cite{Assmann:2004gc}. The luminosities at the various energies are reproduced in Table 2.
%\section{Results and Conclusions}

\begin{figure}[t]
\centerline{\includegraphics[width=8 cm]{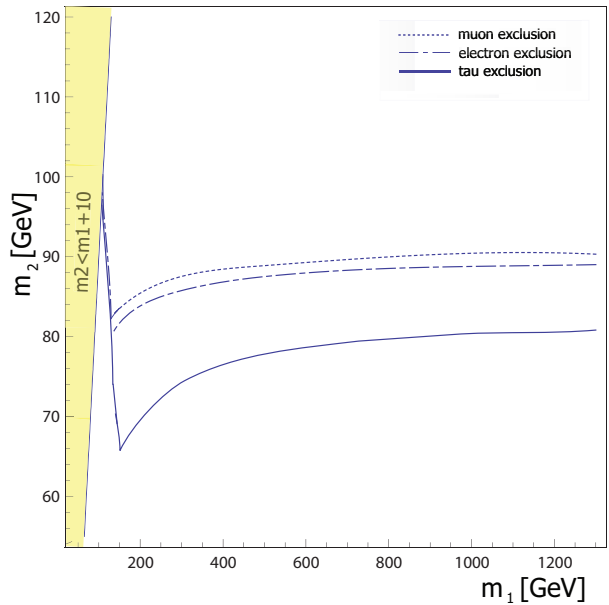}}
\caption{Bounds on $m_1$ and $m_2$ when $N_1$ decays to eW, $\mu$W, and
$\tau$W  respectively.  The region above the lines is allowed. }%%
\label{fig:2pgm}
\end{figure}

The production cross sections for these processes can be analytically calculated to be
\bea
 \sigma_{N_1N_1}=
{1\over 24\pi }{{\left(E_{cm}^2/4-m_{1}^2\right)^{3/2}}\over  E_{cm}}
({g\cos\theta\over \cos\theta_W})^4
\left((-{1\over 2}+\sin^2\theta_W)^2+\sin^4\theta_W\right){ 1\over (E_{cm}^2-m_Z^2)^2}
\\
  \sigma_{N_2N_2}=
{1\over 24\pi }{{\left(E_{cm}^2/4-m_{2}^2\right)^{3/2}}\over  E_{cm}}
({g\cos\theta\over \cos\theta_W})^4
\left((-{1\over 2}+\sin^2\theta_W)^2+\sin^4\theta_W\right){ 1\over (E_{cm}^2-m_Z^2)^2}
 %(E^2-m_{N_1}^2)
\\
 \sigma_{N_1N_2}={1\over 8\pi}{p_1\over  E^3_{cm}}(\cos^2\theta \sin^2\theta)({g\over \cos\theta_W})^4
\left((-{1\over 2}+\sin^2\theta_W)^2+\sin^4\theta_W\right){E_{cm}^2\over (E_{cm}^2-m_Z^2)^2}\nonumber
\\(2E_1E_2+2p^2_1/3+2m_{1}m_{2})
\eea
where in the last line we have defined
\bea
p_1={\sqrt{(E_{cm}^2-(m_{1}+m_{1})^2)(E_{cm}^2-(m_{2}-m_{1})^2)}\over 2E_{cm}}
\\
E_1={(E_{cm}^2-m_{2}^2+m_{1}^2)\over 2E_{cm}}
\qquad\qquad
E_2={(E_{cm}^2+m_{2}^2-m_{1}^2)\over 2E_{cm}}
\eea
(note that $\theta$ in the above formulae is the mixing angle (\ref{mixingangle}), not a kinematic variable.)

We now find the expected number of events and require that they be within the exclusion limits set in \cite{Achard:2001qw}.
This produces exclusion regions in $m_1$-$m_2$ parameter space.
These are shown in Figure 1.

We can understand the features of this plot as follows.
The production cross section $ \sigma_{N_1N_1}$ is suppressed by
$\cos^4\theta$.  In the Majorana limit ${m_2\over m_1}\rightarrow \infty$,
this factor is 1, and we return to the 1-neutrino analysis performed at LEP.
As $m_2$ decreases, $\cos\theta$ decreases, and the production cross section is suppressed.
Eventually when $m_2$ is small enough, the production of $N_1N_2$ and $N_2N_2$ turns on.
and the total cross section again increases.
The total cross section therefore first decreases and then increases as $\cos\theta$ varies
from 1 to ${1\over 2}$; correspondingly the neutrino mass constraints first weaken and then tighten.

For very low mass differences, the decay channel $N_2\rightarrow lW$ may open up, as explained above.
This means that we must include interference effects, which are model dependent since they depend on the unknown mixing angles between the fourth generation and the first three generations. For small mass differences, we may also have new effects like displaced vertices, which we have not considered. For these
reasons, we have excluded from our analysis the region where the mass difference is less than 10 GeV.

By construction,when $m_2$ is much larger than $m_1$, we find the old exclusion limits for
$N_1$. However, when $m_2$  is not very large, the mass bound on $N_1$ is significantly lowered.
These new bounds are shown in Table 3.

\section{Conclusions}

If a fourth generation exists, relatively light righthanded neutrinos must exist in order
to  generate a sufficiently large  neutrino mass. The existence of these extra states modifies
search strategies for the leptonic sector of the fourth generation.
In particular, we have shown that
current LEP searches, which put strong bounds on a single Majorana neutrino,
can be considerably weakened when this more general spectrum is taken into account.
If the lighter neutrino decays to eW  or $\mu$W, the neutrino mass limit
can be reduced to about 80 GeV.
In the case where the fourth generation neutrino
primarily decays to $\tau$W,  we find that these   neutrinos may be as light as  62.1 GeV.

There are several immediate directions for further research.  To complete the study of the leptonic sector, the charged lepton should also be included in the analysis. This already leads to several options for the mass spectra, with possible multilepton signals at colliders.
More generally,
the phenomenology of fourth generation particles with two light neutrinos is a fascinating topic for further searches.
We would expect any heavy fourth generation lepton  to cascade down through the neutrino mass states creating signatures with many final state leptons and missing energy.  If the W-tau-neutrino coupling is dominant, we may have fourth generation pair production with decays to final states with up to 14 final state particles and a large amount of missing energy.  If the fourth generation neutrinos are highly boosted, this raises the possibility of spectacular signals like lepton jets~\cite{ArkaniHamed:2008qp}.

The Tevatron is also capable of searching for the fourth generation leptons directly. In recent work~\cite{Rajaraman:2010wk}, it was shown that the Tevatron can significantly improve the LEP bounds for a Majorana neutrino, with a possibility of excluding neutrinos with mass up to 175 GeV. It would be very interesting to extend this analysis to the two-neutrino case, and obtain the corresponding bounds.

It should also be noted that the Majorana neutrinos decay half the time to same sign leptons (i.e. the decay products are $l^+l^+W^-W^-$). Looking for same sign leptons
significantly reduced the background for the Tevatron search. LEP did not incorporate this event signature in their analysis. A reanalysis of LEP data looking for same sign lepton events also has the potential to significantly improve the reach.

The fourth generation neutrinos can also affect Higgs searches at the Tevatron and LHC. If the neutrino is near the mass limit of 62 GeV, a Higgs with mass in the range between 125 and 160 GeV will primarily decay to these neutrinos, with an  unusual signal of $WW\tau\tau$ (a related analysis has been performed in
\cite{CuhadarDonszelmann:2008jp}).
It would be very interesting to incorporate this decay mode
into  Higgs searches at the Tevatron and LHC. We hope to return to these issues in future work.

\begin{table}[t]
\begin{center}
\begin{tabular}{|c|c|c|}
\hline
$N_1$ Decay Mode &  Previous bounds & New bounds  \\
\hline
W$\tau$ & 80.5 & 62.1 \\
\hline
W$\mu$ & 89.5 & 79.9 \\
\hline
W $e$  &  90.7 & 81.8 \\
\hline
\end{tabular}
\caption{Bounds on $N_1$ mass in GeV  for the various decay channels.}
\label{tab:N1bounds}
\end{center}
\end{table}

\section{Acknowledgements}

We are grateful to T.~Tait and D.~Kirkby for useful discussions.
A.R. is supported in part by NSF Grant No. PHY-0653656.

\end{document}